\documentclass[lettersize,journal]{IEEEtran}
\usepackage[colorlinks,bookmarksopen,bookmarksnumbered,citecolor=red,urlcolor=red]{hyperref}
\usepackage{lipsum}
\usepackage{graphicx,subfig}
\usepackage{mathtools,cuted,float,amsmath,amssymb,amsmath,amsthm,color,epsfig,anyfontsize,t1enc,cite}
\usepackage{mathrsfs,epstopdf}
\usepackage{lipsum} 
\newlength{\subcolumnwidth}

\newcommand{\nextsubcolumn}[1][]{%
	\cr\noalign{\hfill}
	\if\relax\detokenize{#1}\relax\else\hsize=#1\setlength{\subcolumnwidth}{\hsize}\fi
}

\begin{document}

	\title{Multi-hop Sub-THz-FSO: A Technique for High-rate Uninterrupted Backhauling in 6G}
	\author{
		\IEEEauthorblockN{Praveen Kumar Singya$^{1}$, Behrooz Makki$^{2}$, \IEEEmembership{Senior Member, IEEE}, Antonio D'Errico$^{3}$, \IEEEmembership{Senior Member, OPTICA}, and Mohamed-Slim Alouini$^{4}$, \IEEEmembership{Fellow, IEEE}}
		\thanks{$^{1}$P. K. Singya is with the Electrical and Electronics   Engineering (EEE) Department, Atal Bihari Vajpayee-Indian Institute of Information Technology and Management (ABV-IIITM), Gwalior, 474015, India (e-mail:praveens@iiitm.ac.in)}
		\thanks{$^{2}$B. Makki is with Ericsson Research, Ericsson, 41756 Göteborg, Sweden
			(e-mail:behrooz.makki@ericsson.com)}
		\thanks{$^{3}$D'Errico is with Ericsson Research, Ericsson, 56124 Pisa, Italy (e-mail:antonio.d.errico@ericsson.com)}
		\thanks{$^{4}$M.-S. Alouini are with the Computer, Electrical, and Mathematical Science and Engineering (CEMSE) Division, King Abdullah University	of Science and Technology (KAUST), Thuwal 23955-6900, Saudi Arabia (e-mail:slim.alouini@kaust.edu.sa)}}
	\maketitle
	
	\begin{abstract}
		Moving towards $6^{\text{th}}$ generation (6G), backhaul networks require significant improvements to support new use-cases with restricted joint capacity and availability requirements. In this paper, we investigate the potentials and challenges of joint sub-teraHertz (sub-THz) and free space optical (FSO), in short sub-THz-FSO, multi-hop networks as a candidate technology for future backhaul communications. As we show, with a proper deployment, sub-THz-FSO networks have the potential to provide high-rate reliable backhauling, while there are multiple practical challenges to be address before they can be used in large-scale.
	\end{abstract}
	\begin{figure*}[t]
		\centering
		\includegraphics[width=6in,height=4in]{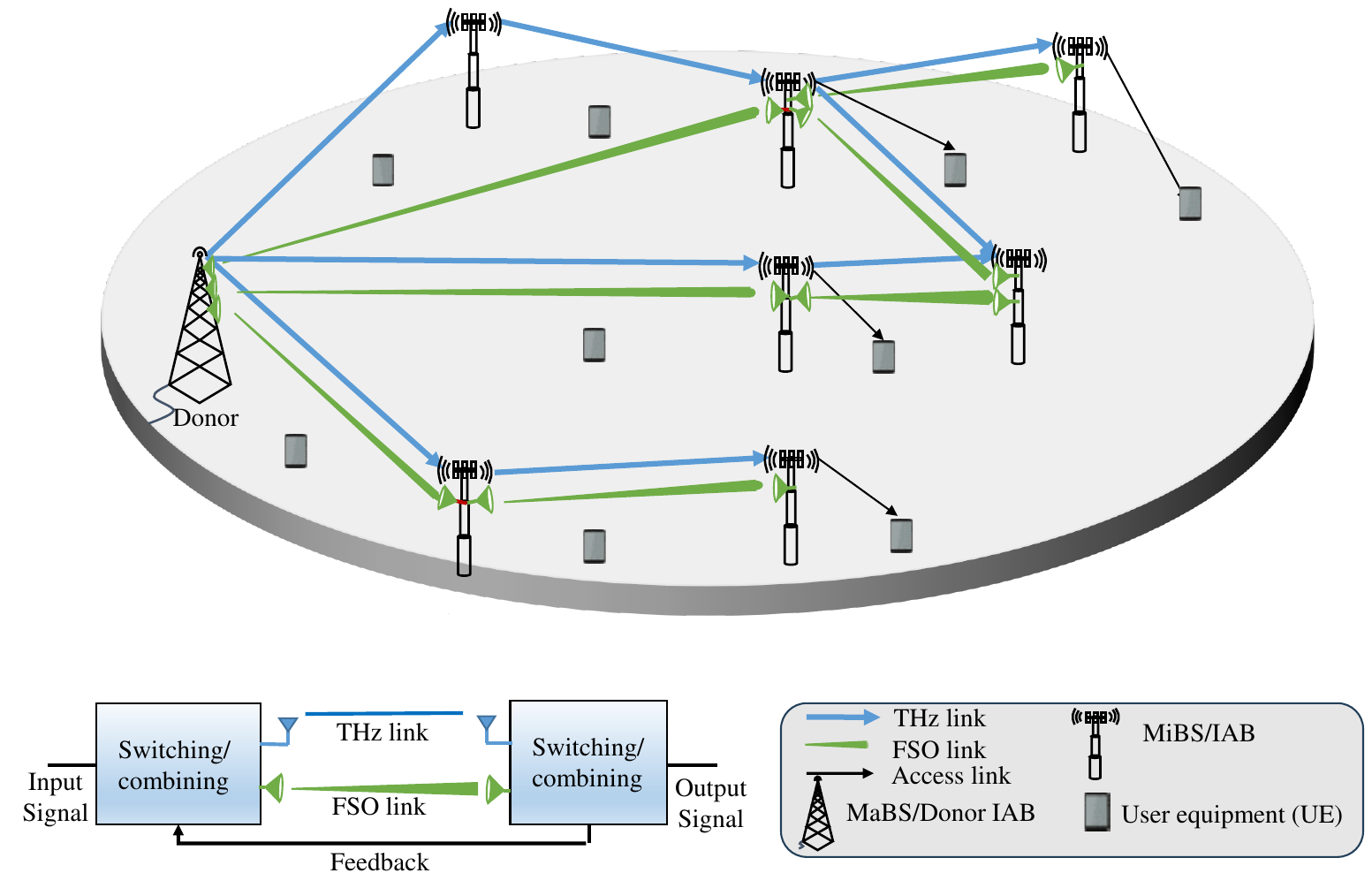}
		\caption{\small{ A generalized multi-hop and mesh IAB network with different types of switching or combination between the links.}}
		\label{Fig1}
		\hrulefill
	\end{figure*}
	\section{Introduction}
	The global mobile data traffic is expected to increase from around 33 exabyte (EB) per month in 2019 to around 164 EB per month in 2025, and the traffic will continue to increase rapidly as we move towards $6^{\text{th}}$ generation (6G) (although it is currently difficult to come up with an accurate number). Such a large traffic needs to be transferred through backhaul networks. Two dominating backhaul technologies are microwave and fiber. In general, the choice between microwave and fiber is not much about the capacity, but rather depends on, e.g., the fiber presence, total cost of ownership (TCO), regional preferences and the connection type; While fiber is of high interest in east Asia, wireless backhaul has a large part of the backhauling market in Europe. Specifically, on a global scale, we foresee a 50/50 split between microwave and fiber for mobile backhaul by 2030 (see \cite{EricOutlooknew} and its previous versions).\par
	Application-wise, fiber is normally used for transport closer to or within the core network, where an aggregated traffic from multiple base stations (BSs) is handled. Microwave, on the other hand, is mainly used as last-mile access as well as the cases where street trenching/digging is expensive or not allowed (for instance, in historical areas). Also, compared to fiber, microwave backhauling has shorter time-to-market and flexibility, at the cost of peak rate.  
	Wireless backhaul currently operates in licensed point-to-point (PtP) spectrum in the range of 4–70/80 GHz. However, W-band (92-114 GHz) and D-band (130-175 GHz), sometimes referred to as sub-THz bands, are considered as the future bands for backhauling in 6G. This is particularly because 26 GHz (24.25-27.5 GHz), 40 GHz (37-43.5 GHz), and 66-71 GHz have been released for access communications and, thereby, usage of backhaul will eventually be transitioned from these bands, to avoid interferences. \par
	As a rule of thumb, higher frequencies are more prone to, e.g., tree foliage, rain  fading, and beam misalignments.
	For instance, considering the mmWave and sub-THz frequencies (30-300 GHz), a 2 mm/hr light rain results in approximately 2.5 dB/km loss, whereas, a heavy rain of  50 mm/hr leads to approximately 20 dB/km loss \cite[Fig. 8]{hemadeh2017millimeter}. 
As a result, high bands may be restricted to the cases with short hops or use-cases with low availability requirements which, indeed, limits their business market. To address such challenges, there are different solutions; Multi-hop communications, following similar or different concept as in the integrated access and backhaul (IAB) specified by 3GPP in Rel-16-18\footnote{3GPP does not specify the IAB networks for W- and D-bands.} \cite{3GPPTS,madapatha2020integrated}, can help to extend the coverage. On the other hand, radio frequency (RF) multi-band operation, i.e., the combination of high and low RF bands increases the capacity/availability \cite{EricOutlook}. Specially, multi-band operation is of interest for the cases with diverse traffic models where for instance the low bands cover the cases with low capacity/high availability requirements and high bands transfer the traffic with high capacity/low availability requirements. 
\par
With the existing backhaul technologies, there is a tradeoff between the capacity and availability, where they can not be achieved simultaneously. This may be problematic in the 6G era where new use-cases such as virtual reality, AI, etc. may be defined with restricted \textit{joint} requirements on capacity and availability. Here, one possible solution  is the combination of sub-THz and free space optical (FSO) technologies with the same order of high capacity but diverse characteristics, so that the diversity boosts the availability at high rates. 

With this background, in this paper, we study the potentials and challenges of sub-THz-FSO multi-hop communication as a candidate technology for high-rate uninterrupted backhauling in 6G. As we show, such combined networks have the potential not only to boost the capacity but also to compensate for the imperfection effects such as pointing error. Moreover, we discuss the key issues which need to be addressed before multi-hop sub-THz-FSO can be used in practical backhaul networks.

\section{High-rate Reliable Backhauling via Multi-hop Sub-THz-FSO Communications }
Although multi-band operation is appropriate for some scenarios, use-cases such as online video streaming, virtual reality, online gaming, online AI-based training, etc., may impose strict joint high capacity and availability requirements on the backhaul networks, which are difficult to achieve via multi-band operation. In such cases, the combination of high-band microwave and FSO, sometimes referred to as sub-THz-FSO, can be a candidate backhauling technology. This is particularly because, as opposed to multi-band operation where the high and low band links have significantly different capacity and availability characteristics, in the sub-THz-FSO, both links provide high capacity. Importantly, these links are complementary because the sub-THz (resp. the FSO) signal is attenuated by rain (resp. fog/clouds), while the FSO (resp. the sub-THz) signal is not. Thus, the joint implementation of sub-THz and FSO can guarantee high-rate reliable backhauling in different weather conditions. Such a solution will be of interest particularly in dense urban areas where, due to the deployment of small cell site/high frequency spectrum, the peak backhaul rate will reach up to 20 Gbps by 2025 \cite{EricOutlook} with >80\% of traffic related to video-based use-cases \cite[Fig. 9]{URL2}
and even higher numbers are foreseen in 6G with the roll out of, e.g., distributive multiple input and multiple output (DMIMO), IAB systems, etc.\par
Both sub-THz and FSO links are prone to hardware impairments such as pointing errors. However, exploiting the diversity and limiting the hop length via multi-hop communications, as illustrated in Fig. \ref{Fig1}, either in the IAB or non-IAB based fashions, can limit the hardware impairment/pointing error effects. Here, multi-hop deployment with short hops is of special interest because both the FSO and the sub-THz links require strong line-of-sight (LOS) links, which may not be available as the hop length increases in dense urban areas. The implementation of the sub-THz and the FSO links can be in parallel or serially. With parallel implementation, e.g., \cite{singyaperformance2022,singya2023high}, both links operate in switching/combining manner to provide diversity. With serial implementation, e.g., \cite{makki2017performance,singya2021performance}, one of the links works as a back-up when the other link collapses. Here, because the sub-THz and FSO links provide the same order of rates, by switching between different links, no big performance drop is experienced, as opposed to the cases with multi-band operation. \par

{\renewcommand{\arraystretch}{1.3}
	\begin{table*}
		\centering 	
			\caption{Considered parameters for the simulations.}
				\begin{tabular}{|ll|ll|ll|}
			\hline
			\multicolumn{2}{|c|}{Sub-THz backhaul link}                                    &\multicolumn{2}{c|}{FSO backhaul link}                                    & \multicolumn{2}{c|}{mmWave access link}                               \\ \hline
			\multicolumn{1}{|l|}{Parameters}                      & Value     &	\multicolumn{1}{l|}{Parameters}                      & Value     & \multicolumn{1}{l|}{Parameters}                      & Value      \\ \hline
			\multicolumn{1}{|l|}{Frequency }       & 119 GHz   & \multicolumn{1}{l|}{Wavelength }       & 1550 nm & \multicolumn{1}{l|}{Frequency }       & 30 GHz     \\ 
			\multicolumn{1}{|l|}{Tx. antenna gain } & 55 dBi   &\multicolumn{1}{l|}{Strong atmospheric turbulence: $C_n^2$} & 1$\times 10^{-12}$m$^{{-2}/{3}}$   & \multicolumn{1}{l|}{Tx. antenna gain} & 40 dBi     \\ 
			\multicolumn{1}{|l|}{Rx. antenna gain}  & 55 dBi    & \multicolumn{1}{l|}{Moderate atmospheric turbulence: $C_n^2$} & 5$\times 10^{-13}$m$^{{-2}/{3}}$    & \multicolumn{1}{l|}{Rx. antenna gain} & 40 dBi     \\ 
			\multicolumn{1}{|l|}{Pressure}      & 101325 Pa & \multicolumn{1}{l|}{Strong atmospheric turbulence parameter: $\alpha_{\text{F}}$,$\beta_{\text{F}}$}      & 4.343, 2.492 & \multicolumn{1}{l|}{Rain attenuation}             & 0 dB/km   \\ 
			\multicolumn{1}{|l|}{Temperature}               & 298 K     & \multicolumn{1}{l|}{Moderate atmospheric turbulence parameter: $\alpha_{\text{F}}$,$\beta_{\text{F}}$ }               & 5.838, 4.249    &  \multicolumn{1}{l|}{Oxygen absorption}   & 15.1 dB/km    \\ 
			\multicolumn{1}{|l|}{Humidity}      & 50$\%$    & \multicolumn{1}{l|}{Opt. to elect. conv. coeff.}      & 1    & \multicolumn{1}{l|}{}                                &          \\ 
			\multicolumn{1}{|l|}{Receiver radius}             &  20 cm    & 	\multicolumn{1}{l|}{Receiver radius}             & 20 cm    & \multicolumn{1}{l|}{}                                &            \\ 
			\multicolumn{1}{|l|}{Beamwidth}             & 50 cm     & \multicolumn{1}{l|}{Beamwidth}             & 40 cm    & \multicolumn{1}{l|}{}                                &            \\ 
			\multicolumn{1}{|l|}{Jitter st. deviation}             &  6 cm    & 	\multicolumn{1}{l|}{Jitter st. deviation}             & 5 cm    & \multicolumn{1}{l|}{}                                &            \\ \hline
		\end{tabular}
		\label{ParametersN}
		\end{table*}
}

			\begin{figure*}[h!]
				\centering
				\includegraphics[width=4.5in,height=3.2in]{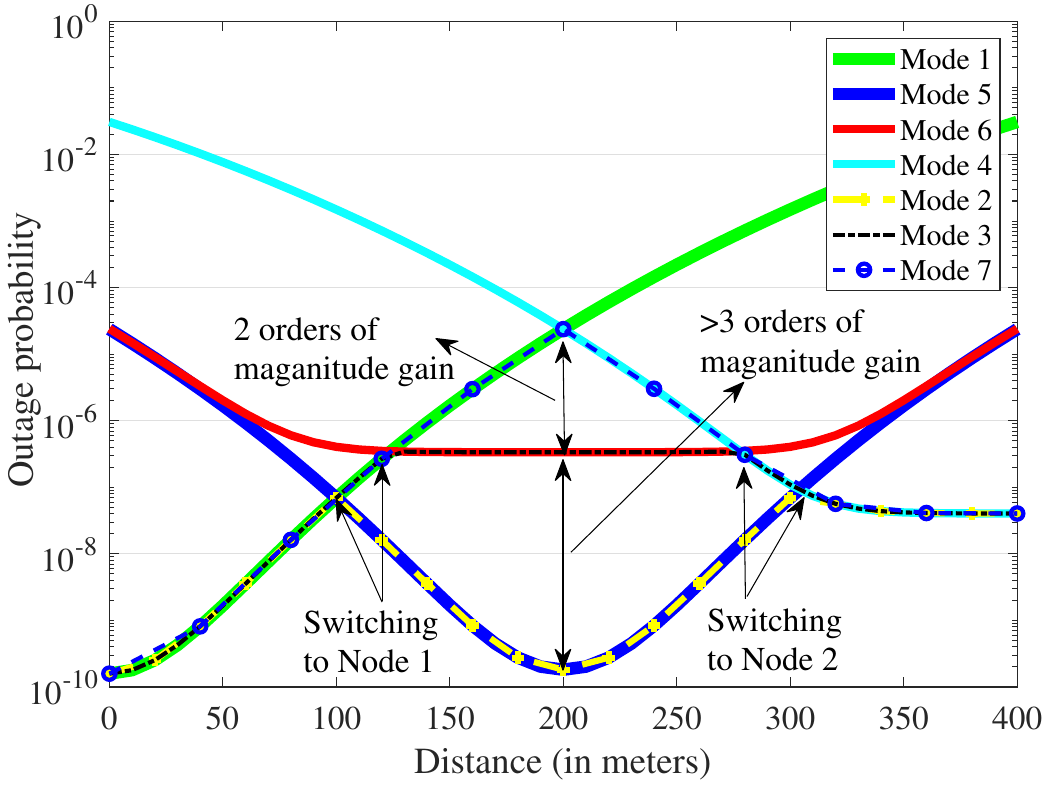}
				\caption{\small{Impact of UE's movement from donor-Node 1-Node 2 on the E2E outage probability. For analysis, we consider moderate atmospheric turbulence, i.e., $C_n^2= 1\times 10^{-12} \text{m}^{-2/3}$}, for the FSO link. For the sub-THz link, the $\alpha-\mu$ fading parameters are $\alpha=2, \mu=3$. Further, we consider 2 antennas at the sub-THz transmitter and receiver. Finally, for the links to the UEs, we assume the mmWave access link's Nakagami-m fading parameter is m=1 with 2 transmit and receive antennas. Transmit power of the FSO and sub-THz links is set to 30 dB. }
				\label{Fig2}
			\end{figure*}

Figure \ref{Fig1} presents general cases with multi-hop and mesh sub-THz-FSO networks. With multi-hop and IAB networks, considering large number of hops affects the system performance significantly due to the increase in end-to-end (E2E) latency and  traffic congestion \cite{madapatha2020integrated}. For this reason, it is probable that in practice the number of hops is limited to a maximum of 2 or 3 hops \cite{madapatha2020integrated}. Also, with a proper network planning, the need for meshed communication may be low and, given the additional complexity of meshed networks, multi-hop networks may suffice, at least in the early stages of 6G \cite{madapatha2021topology}.  \par
Considering an IAB-based setup, each IAB node serves the UEs in its coverage area as well as its child IAB nodes. Further, each IAB node may have different signal-to-noise ratio (SNR) thresholds/spectral efficiency requirements because the SNR threshold is directly proportional to the number of UEs/child IAB nodes served by each IAB node. 
\par

As an illustrative example for the potentials of multi-hop sub-THz-FSO networks, in Fig. \ref{Fig2} we investigate the outage probability of UEs in Fig. \ref{Fig1} when they move between their closest IAB nodes. The considered parameters for various links are summarized in Table \ref{ParametersN}.  Particularly, the results are evaluated by considering Gamma-Gamma distributed atmospheric turbulence for the FSO backhaul link. For pointing error, Rayleigh distribution model is considered with the assumption of identical jitter standard deviation on both axis with zero boresight error. For the multi-antenna sub-THz backhaul link, we consider $\alpha-\mu$ distributed small scale fading and Rayleigh distributed misalignment error. In the access link from the serving node to the UE, the multi-antenna mmWave links are assumed to follow Nakagami-m distribution. In general, the mathematical channel modeling, pointing/misalignment errors, and path-loss modeling of different links follow the same model as in \cite{singya2023high}  while the models are extended to the cases with multiple RF antennas at each node. Also, we consider a rate requirement of  0.1 bps/Hz  per UE and assume 10 UEs per to be served per node. The height of donor/child IAB node is 60 m. 	
In Fig. \ref{Fig2}, we analyze the performance by considering a single UE with different situations: 
\begin{itemize}
	\item Mode 1: The UE is always served directly by the donor node.
	\item Mode 2: The UE is served by a two-hop donor-Node 1-Node 2 backhaul network followed by an access link to the UE, where each node has both sub-THz and FSO links (the lower branch of Fig. 1). The UE moves from donor towards Node 2. Each hop is of length 200m.
	\item Mode 3: The same as Mode 2 but the intermediate node 1 has only sub-THz transceiver and does not have FSO transceiver (the upper branch of Fig. 1). Instead, there is a direct FSO connection from the donor to Node 2. Such a setup enables us to evaluate the effect of parallel links.
	\item Mode 4: The UE is served by a single-hop backhaul donor-Node 2 network followed by an access link to the UE where the backhaul hop length is 400 m. That is, the UE will be served either by the donor or Node 2 placed at 400 m distance of the donor, depending on the channel qualities (the middle branch of Fig. 1).
    \item	Mode 5: Considering the lower branch of Fig. 1, the UE is always served by the middle node (Node 1) which has both sub-THz and FSO links of 200 m in backhaul. Comparison between Mode 2 and Mode 5 highlights the effect of handovers. 
	\item Mode 6: Considering the upper branch of Fig. 1, the UE is always served by the middle node (Node 1) which has only sub-THz transceiver in backhaul.
		\item Mode 7: Considering the middle branch of Fig. 1, the UE is always served by Node 2 which has 400 m sub-THz and FSO links in backhaul.
\end{itemize}
\par

According to the figure, in Mode 1, i.e., the cases without intermediate IAB nodes, the outage probability of the UE increases rapidly as the UE moves away from the donor node, until it reaches 10$^{-2}$ at 400 m distance. Similar trend is observed in Mode 7 when the UE moves away from its serving node. Also, in Mode 4 the worst performance is observed at around 200 m when the UE is in between the two serving nodes. These indicate the necessity of the intermediate nodes/short backhaul links, to guarantee an almost uniform performance in the coverage area.  \par

In both Modes 2 and 3, as the UE moves away from the donor node, it is first connected directly to the donor, then switches to the middle node and finally handovers to the last node. Thus, the minimum outage performance is observed when the UE is close to the middle node. However, compared to Modes 1 and 4, in Modes 2-3, more uniform performance is observed in the whole coverage area, and the outage probability always remains in the acceptable range. For example, at 200 m, Mode 2 and Mode 3 provide respectively around 3 and 2 orders of magnitude improvement in the outage probability, compared to Mode 4  which has no middle node.\par

In all considered ranges, better performance is observed in Mode 2, compared to Mode 3, i.e., when the middle node is also equipped with both the sub-THz and the FSO transceivers. For example, at 200 m, Mode 2 provides around 3 orders of magnitude  improved performance over Mode 3. This is intuitively due to the lack of FSO transceiver in the middle node which affects the availability of the backhaul link. 


\begin{figure}[t]
	\centering
	\includegraphics[width=3.5in,height=2.7in]{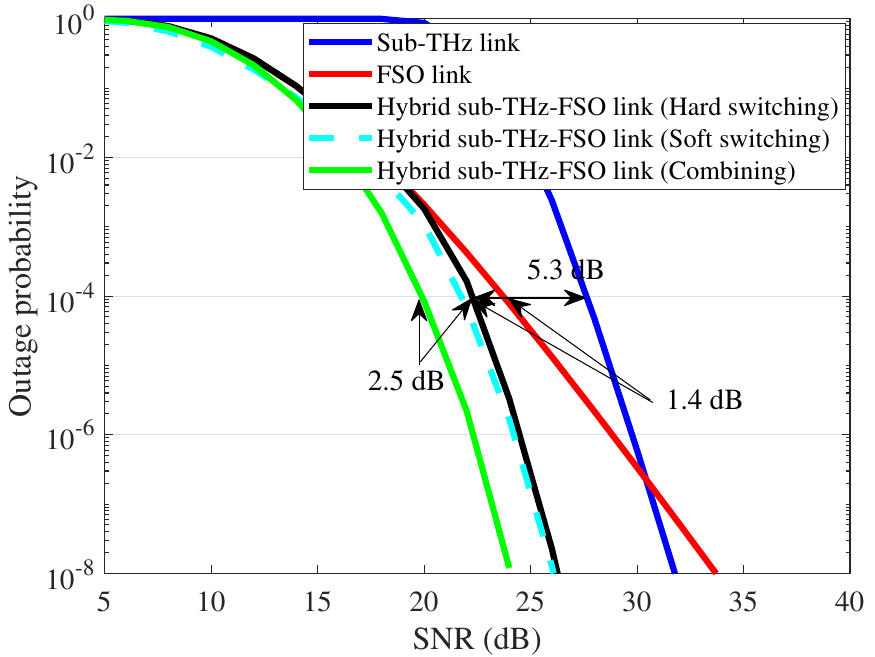}
	\caption{\small{Outage probability versus the transmit SNR of various links. For FSO link, moderate atmospheric turbulence, i.e., $C_n^2= 5\times 10^{-13} \text{m}^{-2/3}$ is considered. For the sub-THz link, we consider $\alpha=2, \mu=3$ in the $\alpha-\mu$ model, and 2 transmit and receiver antennas are considered. For soft switching, upper and lower FSO SNR thresholds are 6 dB and 4 dB, respectively (see \cite{singyaperformance2022} for details of the soft switching method). The sub-THz SNR threshold is set to 5 dB.}}
	\label{Out_comp}		
\end{figure}

To further demonstrate the benefits of the parallel implementation of the sub-THz and FSO links, Fig. \ref{Out_comp} shows the comparison of outage probabilities of various links in single hop of 200 m length. For comparison, we consider FSO link, sub-THz link, and hybrid sub-THz-FSO link with hard switching, soft switching, and combining  methods at the receiver. With combining, both the sub-THz and the FSO signals are combined for decoding using maximum ratio combining (MRC). With hard switching, only the signal of the link with higher SNR is used for decoding. With soft switching, on the other hand, rather than having one SNR threshold, dual FSO SNR thresholds are used, which allows the FSO link to be active for  longer duration and reduces the frequent switching between the sub-THz and FSO links. That is, with soft switching, a link operating in, e.g., FSO mode may still continue with FSO-based communication even if the SNR of the FSO link drops slightly below the SNR observed in the sub-THz link, or vice versa. In this way, with soft switching, we switch to the other link only if a deep performance drop is observed or the active link experiences poor condition for a long time (see \cite{singyaperformance2022} for the details of hard and soft switching methods). \par

From Fig. \ref{Out_comp}, we observe that for the considered set of parameters, the FSO link outperforms the sub-THz links in low and medium SNR regimes, however, the sub-THz link outperforms the FSO link at high SNRs. Considering different switching or combining methods, the parallel implementation of the links outperforms the performance, compared to the cases using only one of the FSO or sub-THz links. This is mainly because of the additional diversity of the hybrid link. For instance, with hard switching and an outage probability of 10$^{-4}$, the hybrid sub-THz-FSO link results in 5.3 and 1.4 dB gain over the sub-THz and FSO links, respectively. As expected, MRC-based combination of the signals improves the performance, compared to the switching-based methods, at the cost of complexity. For instance, at outage probability 10$^{-4}$, 2.5 dB gain is achieved via the combination, compared to hard switching method. Finally, for all considered ranges of SNR, almost the same performance is observed in the soft and hard switching methods. However, soft switching,
 reduces the back-and-forth switching between the links significantly. This is due to the fact that, with soft switching, one of the links remain active for longer duration. Conclusively, in practice, the soft switching has superior performance, compared to the hard switching.\par
 \begin{figure}[h]
 	\centering
 	\includegraphics[width=3.5in,height=2.8in]{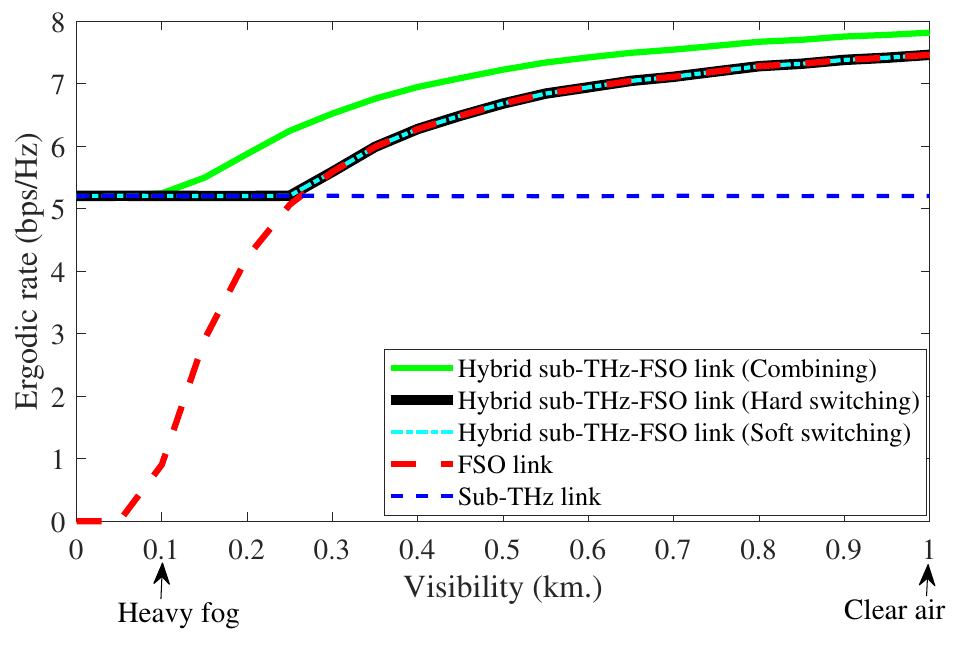}
 	\caption{\small{Ergodic rates of various links versus the visibility of the FSO link. For the FSO link, strong atmospheric turbulence, i.e., $C_n^2= 1\times 10^{-12} \text{m}^{-2/3}$ is considered. For the sub-THz link, $\alpha=2, \mu=3$, and 2 transmit and receiver antennas are considered. Transmit power of both links is set to 30 dB. For soft switching, upper and lower FSO SNR thresholds are 6 dB and 4 dB, respectively. The sub-THz SNR threshold is set to 5 dB.}}
 	\label{Cap_Comp}		
 \end{figure}
Figure \ref{Cap_Comp} demonstrates the ergodic rates of the individual and hybrid links for different visibility conditions of the FSO link. This is due to the fact that heavy fog reduces the visibility, which affects the performance of the FSO link  considerably. As shown, the FSO link’s
performance is quite poor at low visibility while, as the visibility increases, the FSO link’s performance surpasses the sub-THz link’s performance. In this way, with both switching and combination methods, the hybrid sub-THz-FSO link guarantees high and almost uniform rate in different channel conditions.

\begin{figure}[t]
	\centering
	\includegraphics[width=3.5in,height=2.7in]{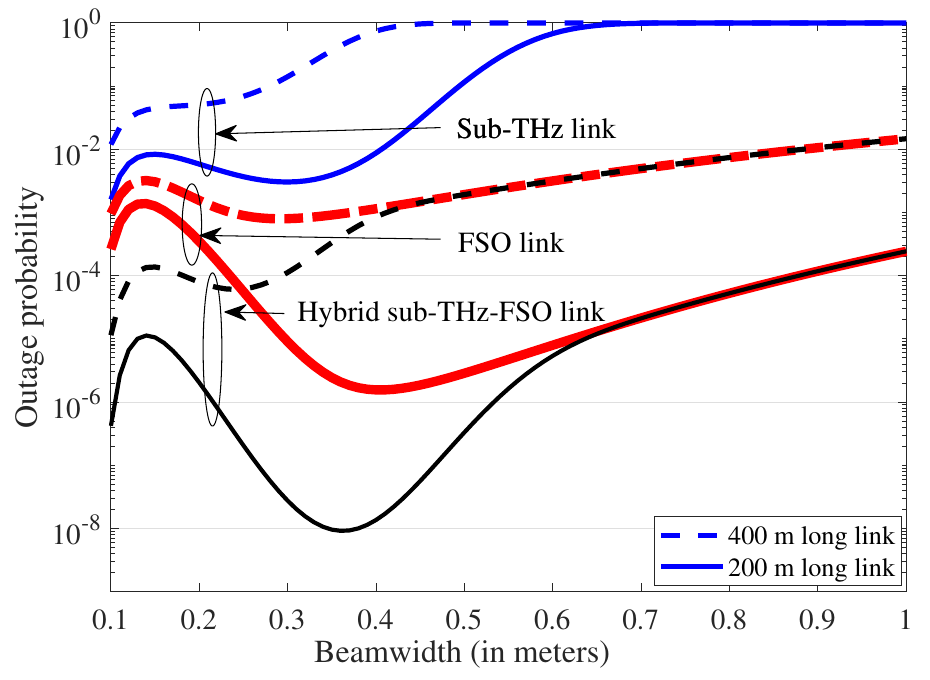}
	\caption{\small{Outage probability versus beamwidth to show the impact of pointing error on various links. For the FSO link, moderate atmospheric turbulence, i.e., $C_n^2= 5\times 10^{-13} \text{m}^{-2/3}$ is considered. For the sub-THz link, $\alpha=2, \mu=3$, and 2 transmit and receiver antennas are considered. Transmit power of both links is set to 30 dB.}}
	\label{Fig4}		
\end{figure}

In practice, both the FSO and sub-THz links are subjected to pointing error, particularly as the hop length increases. However, the diversity offered by the hybrid link can partly compensate for the pointing errors/beam misalignment. To highlight this point,
in Fig. \ref{Fig4}, the outage probability of various links is illustrated versus the beamwidth. We consider the same receiver radius 0.1 m and same jitter standard deviation 0.12 m for both the FSO and the sub-THz links. Also, the pointing error follows the Rayleigh fading model.\par
 As demonstrated, the sub-THz link suffers significantly from beam misalignment, specially as the hop length increases. Particularly, both the 400 m and 200 m sub-THz links reach in complete outage after 0.4 m and 0.6 m beamwidth, respectively. On the other hand, the FSO link is less sensitive to pointing errors. Then, the parallel implementation of the links increases the robustness to pointing error significantly, particularly at low/moderate beamwidths where the outage probability is decreased by at least an order of magnitude, compared to the cases using only one of the links. Also, for all considered individual and hybrid links the minimum outage probability is observed at a finite value of the beamwidth. This is intuitively due to the fact that with further increase in beamwidth less power is collected at the receiver. Conclusively, for a constant jitter standard deviation, each link (sub-THz or FSO) has its optimum performance at a certain beamwidth. Further, increase in beamwidth will further reduce the performance.

\par
\section {Towards Sub-THz-FSO based Backhauling; Challenges and Testbed Evaluations}
Few testbed evaluations on RF-FSO-based backhaul networks have been performed: 
	\begin{itemize}
		\item In \cite{song2022demonstration}, a hybrid FSO/RF-based experimental setup is demonstrated, where the FSO link operates at 1550 nm wavelength and the RF link operates at 13 GHz frequency. From experiments, 10 Gbps data rate through the FSO link with intensity modulation and direct detection whereas the RF link can provide upto 1 Gbps data rates with 1024-quadrature amplitude modulation (QAM) for 4 m distance in an indoor environment.
		
		\item  In \cite{URL3}, an RF-FSO-based backhaul network is developed where the FSO link is capable of providing around 20 Gbps data rates for nearly 20 km in LOS clear weather condition. The FSO link operates at 1550 nm wavelength and the RF link operates at 6 GHz frequency. The RF link is deployed as a backup link to the FSO.
		
		\item In \cite{rodiger2020demonstration}, a hybrid RF-FSO experimental setup is demonstrated where a 300 m link is established. In the proposed configuration, the FSO link provides around 610 Mbps throughput whereas the backup RF link provides a nominal 5.8 Mbps throughput.
				
		\item  In \cite{czegledi2020demonstrating}, a 1.5 km  LOS RF link is established. The link operates at 73.5 GHz frequency with 2.5 GHz channel bandwidth and $8\times 8$ LOS MIMO antenna configuration. The setup achieves 108 Gbps, 125.7 Gbps 139 Gbps throughput with 64 QAM, 128 QAM, and 256 QAM modulation formats, respectively, with 55.6 bps/Hz spectral efficiency.
	\end{itemize}
To the best of our knowledge, there is no testbed for the sub-THz-FSO link evaluation, while \cite{song2022demonstration,URL3,rodiger2020demonstration} consider low RF bands and \cite{czegledi2020demonstrating} does not consider the FSO links. However, the testbed results of the high-band RF links in \cite{czegledi2020demonstrating} indicate that at high frequencies the sub-THz and the FSO links can provide the same order of rates, which can satisfy the joint rate/availability requirements of 6G. \par

Although the evaluations show great potential for sub-THz-FSO links, there are still multiple issues to be addressed before it can be deployed in large-scale:
\begin{itemize}
	\item While  microwave backhauling is a well-established technology implemented globally for decades, FSO is not as mature as microwave backhauling and has been limited mainly to special testbeds/academic investigations. For these reasons, finding a large business market for the joint sub-THz-FSO based backhauling is not trivial.
	
	\item The choice of a proper backhaul technology depends much on the network TCO. For this reason, deep cost analysis of the sub-THz-FSO links, and comparison with alternative technologies is required. 
	\item Microwave backhaul mainly operates on licensed bands. Also, some aspects of microwave backhauling are standardized (although a large part of microwave backhauling is based on proprietary, i.e., non-standardized, solutions). Particularly, with possible extension of the access communication spectrum to higher bands in 6G, the IAB standardization may be revisited to cover high bands. As opposed, FSO is a pure proprietary-based technology operating in unlicensed bands. Then, the combination of these links may require standardization efforts. 
	\item	Different from the microwave links, the FSO links may require regular calibrations and maintenance. Here, automatic calibration via \textit{zero-touch}  maintenance, i.e., with no need of manual work, should be implemented to guarantee robust operation of the link. This is, indeed, an extra cost to be considered in the TCO calculations.     
	\item	As opposed to FSO links which are extremely secure, the microwave links face with security challenges. Hence, to guarantee high E2E security in the sub-THz-FSO links, one needs to boost the security of the sub-THz links.
\end{itemize}

\section{Conclusions}

We studied the potentials and challenges of multi-hop sub-THz-FSO networks as a candidate technology for high-rate reliable backhauling in 6G. As we showed, the diversity and the short hop lengths of these networks compensate for different imperfections and improve the network availability at high rates. However, this is still a niche technology, and there are different practical issues to be addressed before it can be used in large scale.

\bibliographystyle{IEEEtran}
\tiny
\bibliography{Ref_Mag}

	\begin{IEEEbiography}[{\includegraphics[width=1in,height=1.25in,clip,keepaspectratio]{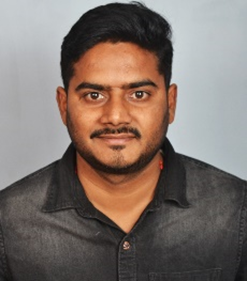}}]{Praveen K. Singya}  is currently an Assistant Professor in ABV-IIITM Gwalior. His research interest includes Integrated Access and Backhaul, FSO, and THz Communication.
\end{IEEEbiography}

\begin{IEEEbiography}[{\includegraphics[width=1in,height=1.25in,clip,keepaspectratio]{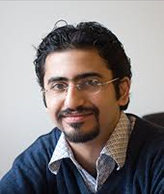}}]{Behrooz Makki} is a Senior Researcher in Ericsson Research, Sweden. His research interests include wireless backhaul, integrated access and backhaul, and intelligent reflecting surfaces. 
\end{IEEEbiography}

\begin{IEEEbiography}[{\includegraphics[width=1in,height=1.25in,clip,keepaspectratio]{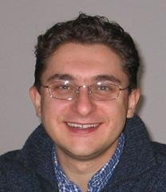}}]{Antonio D’Errico}   is a Senior Researcher and is now a Specialist in photonics and telecommunication systems in Ericsson Research. His research interests include advanced technological solutions for optical networks. He is currently working on photonic-enabling technologies for 6G Mobile Networks.
\end{IEEEbiography}

\begin{IEEEbiography}[{\includegraphics[width=1in,height=1.25in,clip,keepaspectratio]{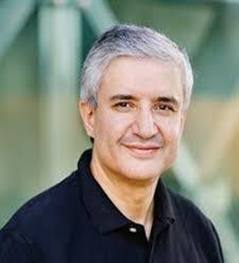}}]{Mohamed-Slim Alouini} is the  Distinguished Professor in the Electrical and Computer Engineering, at KAUST, Saudi Arabia. His research interests include modeling, design, and performance analysis of wireless communication systems.
\end{IEEEbiography}

\end{document}